\documentclass[aps,prd,twocolumn,groupedaddress,nofootinbib,showpacs]{revtex4}
\usepackage{amsmath,amssymb,epsfig}
\usepackage[dvipdfm]{hyperref} 

\newcommand{\rf}[1]{(\ref{#1})}
\newcommand{\beq}{\begin{equation}}

\newcommand{\eeq}{\end{equation}}
\newcommand{\be}{\begin{equation}}
\newcommand{\ee}{\end{equation}}
\newcommand{\bea}{\begin{eqnarray}}
\newcommand{\eea}{\end{eqnarray}}
\newcommand{\eq}[1]{Eq.~(\ref{#1})}
\newcommand{\non}{\nonumber \\*}
\newcommand{\ie}{{i.e.}\ }

\newcommand{\e}{\mbox{e}}
\renewcommand{\d}{{d}}

\newcommand{\m}{\mu}
\newcommand{\K}{K}

\newcommand{\eps}{\epsilon}
\newcommand{\om}{\omega}
\newcommand{\del}{\delta}

\newcommand{\oh}{\frac{1}{2}}

\newcommand{\tr}{\mathrm{tr}\,}

\newcommand{\prt}{\partial}

\newcommand{\cD}{{\cal D}}

\newcommand{\nn}{\nonumber}

\def\gtrsim{\mathrel{\mathpalette\fun >}}
\def\fun#1#2{\lower3.6pt\vbox{\baselineskip0pt\lineskip.9pt
\ialign{$\mathsurround=0pt#1\hfil##\hfil$\crcr#2\crcr\sim\crcr}}}

\def\ga{\gtrsim} 
\begin{document}

\preprint{ITEP--TH--48/13}

\title{Effective QCD string beyond the Nambu--Goto action}

\author
{J. Ambj\o rn$\,^{a,b}$, Y. Makeenko$\,^{a,c}$
and A. Sedrakyan$\,^{a,d}$}

\affiliation{\vspace*{2mm}
${}^a$\/The Niels Bohr Institute, Copenhagen University,
Blegdamsvej 17, DK-2100 Copenhagen, Denmark\\
${}^b$\/IMAPP, Radboud University, Heyendaalseweg 135,
6525 AJ, Nijmegen, The Netherlands\\
${}^c$\/Institute of Theoretical and Experimental Physics,
B. Cheremushkinskaya 25, 117218 Moscow, Russia\\
${}^d$\/Yerevan Physics Institute,
Br.\ Alikhanyan str 2, Yerevan-36, Armenia \\ 
\vspace*{1mm}
{email: ambjorn@nbi.dk \ makeenko@nbi.dk \ sedrak@nbi.dk}
}

\begin{abstract}
We consider the QCD string as an effective string, whose action
describes long-range stringy fluctuations. The leading infrared
contribution to the ground state energy is given by the 
Alvarez--Arvis formula, usually derived using the Nambu-Goto action.
Here we rederive it by a saddle point calculation 
using the Polyakov formulation of the free string, where the 
world sheet metric and the target space coordinates are treated 
as independent variables.   
The next-order relevant in the infrared term in the effective action  is the 
extrinsic curvature term. We show that the spectrum does 
not change order by order in the inverse string length, 
but may change at intermediate distances.

\end{abstract}


\maketitle

\section{Introduction\label{intro}}

As is well understood by now, QCD string is made out of fluxes of the gluon
field. Effective stringy degrees of freedom make sense at the distances larger 
than the confinement scale -- at shorter distances  
quark-gluon degrees of freedom are 
more appropriate owing to asymptotic freedom. The corresponding effective 
theory of long strings can be constructed \cite{PS91} order by order in
the inverse string length and can be consistently quantized in any
number of space-time dimensions. Various approaches to such an effective string 
theory have recently attracted considerable attention 
\cite{Aharony,HD,HD2,HD3,Tep09,Pepe,ABT10,AF11,Pepe2,Mak11a,Casele,Mak11b,ABT11,AFK12,Cas2,Dub,Bahns,Vyas,Dub13,AK13,AT13,Cas3,Bra13}.

The analytical results concern mostly the simplest bosonic
string given by the Nambu--Goto area action or the Polyakov action. 
However, such an action can be only serve as an  approximation 
to the full effective action for the QCD string
at very large distances. With decreasing the distance terms less 
dominant in the infrared will become essential and 
are needed  in order to avoid in particular the tachyonic instability 
of the Nambu--Goto energy spectrum. 

The investigation of the effective string theory is therefore twofold:
firstly, to quantize the pure bosonic Nambu--Goto long string and
to compute its spectrum and,
secondly, to construct the next relevant operators and to compute their
contribution to the spectrum.
We shall deal in this Paper with both of these tasks.

In Sect.~\ref{s:2} we concentrate on the spectrum of the Nambu--Goto 
effective string, using its formulation {\it \'a la} 
Polyakov \cite{Polyakov-1981}. We fix the conformal gauge  
$g_{ab}= \e^{\phi}\delta_{ab}$ and treat
the target-space string coordinates and the world sheet metric as independent.
In a Wilson loop setup they couple via the boundary conditions and we will 
show how one can reproduce the Alvarez--Arvis stringy 
spectrum~\cite{Alv81,Arv83} by a saddle point calculation. 


The essential feature of the Liouville action%
\footnote{Here $d$ is the central charge  of matter fields living on a string 
world sheet. For bosonic
string it is just the dimension of space-time. For fermions it counts 
the number of Dirac particles $2^{[d/2]}$,
each of which has the central charge 1.}
\bea
\label{Liouville}
S_{g}= -\frac{d-26}{96 \pi} \int \d^2 z
\Big[(\partial_a \phi)^2 +\mu^2 \e^{\phi}\Big]
\eea
is the fact that it depends only on the metric, \ie
internal geometry of the surface, 
and does not  depend on external geometry of its embedding into $d$-dimensional
space. This is a quite strong restriction on string properties, which is 
dictated by the original bare string action
\cite{Polyakov-1981}.  However, for other strings, for example for 
the Green--Schwartz superstring,
where the fermions belong to the spinor representation of the group $SO(d)$ of 
the target space,
the effective action of strings depends essentially on external geometry of 
the string world sheet \cite{Kavalov-1986, Kostov-1986, Stora-1987}
due to quantum fluctuations of fermions. Namely, the effective action depends 
now on the second quadratic form $h_{ab}$
of the embedded surface. The simplest model with an inclusion of 
external geometry -- the extrinsic curvature -- was considered in 
\cite{Pol86,Klei86}.
In general, one can expect that similar dynamics of external geometry
of embedded surfaces may appear also in an effective theory of QCD string. 

In Sect.~\ref{s:3} we consider the bosonic string with extrinsic curvature.
We briefly review the standard results 
\cite{OY86,Bra86,GK89,Ger91,PY92} for the ground-state energy 
in the world sheet parametrization and
emphasize that deviations from the spectrum of
the Nambu--Goto string are explicitly seen,
when the coefficient in front of the extrinsic-curvature term 
in the action becomes large. The string then becomes rigid.
We reproduce the same results in the language of the effective
string, using the upper half-plane parametrization. This now
happens because the conformal anomaly changes.
Finally, in Sect.~\ref{s:33} we derive the most general expression for the 
conformal anomaly, accounting for its possible dependence on 
external geometry, and discuss the effect of these additional terms 
on the spectrum.

\section{The effective Nambu--Goto string\label{s:2}}

\subsection{Effective string: Polyakov's formulation}

In the Polyakov string formulation the coordinates
$X_\mu(x,y)$ of string in the target space and the  metric $g_{ab}(x,y)$ 
of the world sheet spanned by the string
are independent. The Polyakov path integral is thus \cite{Polyakov-1981}:
\beq\label{ja1}
W(C) = \int \cD g_{ab} \int \cD X_\mu \; \e^{-
\frac{1}{4\pi\alpha'} \int d^2 z \,\sqrt{g}g^{ab} \partial_a X_{\mu} \cdot
\partial_b X_{\mu}},
\eeq
where $C$ denotes a planar curve in $R^d$
(the Wilson loop in the QCD terminology). It is important to 
recall that the (regularized) functional integration measure $\cD X_\mu$
depends on the metric $g_{ab}$, as the cutoff does for reparametrization 
invariance. 
This fact
 can be viewed as the origin of the conformal anomaly of the theory.


It is convenient to make a partial gauge fixing of 
$g_{ab}(x,y)$ to the conformal gauge $g_{ab}=\e^{\phi}\delta_{ab}$, 
where the scalar (Ricci) curvature $R=-\Delta \phi = -\e^{-\phi} \prt^2 \phi$. 
This can be done globally when the topology of 
the world sheet is that of a disk,
which is the situation we consider here. Thus the setup is 
the following: we have a region $D$ in the complex plane with the 
topology of the disk. It has a boundary $\prt D$. The coordinates
$X_\mu(z)$, $\mu =1,\ldots,d$ represent a map from $D$ to $R^d$.
The Wilson loop $C$ is a $R\times T$ rectangle in $R^d$. We assume it
is located in the (1,2) plane of $R^d$.  $X_\mu(z)$ maps the 
boundary $\prt D$ onto $C$. 

Let $C$ be given by the  
parametrization $x_\mu(s)$, $s$ denoting the ach-length from 
a point $x_\mu(s_0)$,  and assume that $z_0\in \prt D$ is mapped to 
$x_\mu(s_0)$. As is described in \cite{DOP84,Alv83}, the natural coupling
between the length scale of the Wilson loop $C$ in $R^d$ and
the length scale set by the intrinsic metric at the 
boundary  takes place at the boundary by insisting 
that $\dot{X}_\mu^2(z) \propto \e^{\phi(z)}$, 
$z\in \prt D$. Here  $\dot{X}_\mu(z)$ denotes $d X_\mu(z)/dz$ 
along the boundary.
Denote the length of $\prt D$ calculated using the boundary 
metric $e(z) = \e^{\phi(z)/2}$ by $L_\phi$ and 
the length of the Wilson loop in $R^d$ (calculated using $X_\mu(z)$)
by $L$ ($L=2(T+R)$). Then we have 
for a given $\phi(z)$ the boundary condition for $X_\mu(z)$:
\beq\label{ja3}
X_\m(z) = x_\mu(s(\phi;z)),~~~
s(\phi;z)= \frac{L}{L_\phi} \int_{z_0}^z dz\; \e^{\phi(z)/2}
\eeq
for $z\in \prt D$.
Without loss of generality we can choose the constant 
of proportionality between  $\dot{X}_\mu^2(z)$ and  $\e^{\phi(z)}$ to be 
one, i.e.\ $L_\phi = L$.       
Let us introduce the notation $\psi(z) = \phi(z)$ for $z\in \prt D$. 
$\psi(z)$ is a field which lives on the boundary $\prt D$ and a change 
of $\psi(z)$ corresponds to a reparametrization of the boundary.

For a given choice of $\psi(z)$ the equations of motion for $X_\mu(z)$ 
with corresponding boundary equations become
\bea
\prt^2 X_\mu(z) &=&0, \quad \hspace*{15mm} z\in D, \non
X_\mu(z) &= &x_\mu(s(\psi;z)), \quad z\in \prt D.
\label{ja4}
\eea
Denote the corresponding solution by $X_\m^\psi(z)$.
Similarly, the equations of motion for $\phi$ leads to 
the condition that the induced metric 
\be
(g_{\rm cl})_{ab} \equiv \prt_a X^\psi \cdot \prt_b X^\psi
\label{gcla}
\ee
is conformal (or isothermal)
\beq\label{ja4a}
(g_{\rm cl})_{ab} = \oh \del_{ab} \sum_c (g_{\rm cl})_{cc}.
\eeq
If Eqs.\ \rf{ja4} and \rf{ja4a} are satisfied, the surface 
$X_\mu$ is the surface of minimal area corresponding to the curve $C$.

We can now decompose $X_\mu(z)$ as 
\beq\label{ja5a}
X_\mu (z) = X_\mu^\psi(z) + \del X_\mu(z),~~~~\del X_\mu (z) =0~~~z\in \prt D.
\eeq  
The path integral over $X_\mu(z)$ in \rf{ja1} can be performed  
be shifting the  integral to $\del X_\mu(z)$ and is 
given by the conformal anomaly \cite{Polyakov-1981}. The 
end result%
\footnote{We have not written here the boundary terms 
found in \cite{Olesen,Alv83}. They will be presented below (see \eq{Gilkey}).}
 is \cite{DOP84}:
\bea\label{ja5}
W(C) &=& \int \cD \psi(z) \e^{-\frac{1}{4\pi\alpha'} \int d^2 z \, 
\partial_a X^{\psi}  \cdot
\partial_a X^{\psi}}  \non && \times\int \cD \phi(z) \;
\e^{\frac{d-26}{96\pi} \int d^2 z \,
\partial_a \phi\partial_a \phi\,},
\eea
where the functional integral over the Liouville field $\phi(z)$, $z\in D$, is 
performed with the boundary condition $\phi(z) = \psi(z)$, $z\in \prt D$.
It is important to remember that  $\cD \phi(z)$ 
is not the measure of a free field $\phi(z)$ since a 
reparametrization-invariant cutoff is needed in the functional integral, and this 
cutoff will then  depend on $\phi$ 
itself, since $\phi$ is part of the metric. 
 
The Liouville action with the factor $d$ appears as a result of 
the integration over the $\del X_\mu$ and reflects that the 
measure $\cD (\del X_\mu)$ is not invariant under conformal 
transformations of the intrinsic metric $g_{ab}(z) = \e^{\phi(z)}\del_{ab}$. 
The  factor --26 appears as a result of the 
partial gauge fixing of $\cD g_{ab}$ to conformal gauge, leaving 
only the conformal mode $\phi(z)$ to be integrated over, the 
$\phi(z)$ appearing in the path integral in \rf{ja5} 
with boundary condition $\phi(z) = \psi(z)$, $z \in \prt D$.


It is custom to represent the disk amplitude \rf{ja5} as
\be
W(C)=\e^{-S_{\rm eff}(C)},
\ee
introducing an effective action $S_{\rm eff}(C)$.
It is called effective since quantum fluctuations of $X_\mu$ are
already taken into account by the presence of the conformal anomaly.
As we shall see below, the upper half-plane parametrization is
most convenient at this point. 

The field $\phi$ (often called the Liouville field) is generically quantum,
but for long strings and/or large $|d|$ it freezes near the value, minimizing an
effective action. We then have 
\be 
\left\langle g_{ab} \right\rangle _\phi = \rho \delta_{ab}, 
\label{mmm}
\ee
where $\rho$ coincides with the classical induced Weyl factor $\rho_{\rm cl}$
only when $\alpha'\to0$. 
The explicit formulas for $\rho$ and $\rho_{\rm cl}$ will be shortly
presented.


The difference from the Polchinski--Strominger effective string theory
\cite{PS91}, where $g_{ab}$ equals the induced metric at the outset, is that
we now determine $\left\langle g_{ab} \right\rangle _\phi$ by the minimization.

\subsection{UHP parametrization\label{ss:2.2}}

Since our Wilson loop $C$ has been chosen as a $R\times T$ rectangle 
(which we can view as lying in the complex $\om$-plane), an obvious 
solution to Eq.\ \rf{ja4} is to choose $D$ as a $\om_R \times \om_T$
rectangle, and  
\beq\label{Xcl}
X^\psi_1  = \frac{R}{\omega_R} \Re\, \omega, \qquad
X^\psi_2  = \frac{T}{\omega_T} \Im\, \omega,
\eeq
while the rest of the $X_\mu$'s equal 0. This is called the 
world sheet parametrization. However, we are interested in 
using instead the upper half-plane (UHP). The reason is that 
we want to perform the boundary path integral in \rf{ja5} and 
this is best done using the upper half-plane parametrization.

For the upper half-plane parametrization
$z=x+ i y$  with $y\geq 0$ and the boundary is at the real axis ($y=0$)
to be parametrized by $s$.
The  Schwarz--Christoffel map of the upper half-plane onto a rectangle is 
\bea
\omega(z)&=& \sqrt{s_{42}s_{31}}\int_{s_2}^z \frac{\d x } 
{\sqrt{(s_4-x)(s_3-x)(x-s_2)(x-s_1)}}\non
&=&2 
F \left( \sqrt{\frac{s_{31}(z-s_2)}{s_{32}(z-s_1)}},
\sqrt{\frac{s_{32} s_{41}}{s_{42} s_{31}}} \right),
\label{SCmap}
\eea
where $s_1<s_2<s_3<s_4$ ($s_{ij}=s_i-s_j$) are associated with the four corners
of the rectangle.
In \eq{SCmap} $F$ is the incomplete elliptic integral of the first kind and
the normalization factor is introduced for latter convenience.

The new variable $\omega$ takes values inside a  $\omega_R\times \omega_T$ 
rectangle, which now can be given  the meaning
of a world sheet parametrization. 
From \eq{SCmap} we have
\be
\omega_R=2 K\left(\sqrt{1-r}\right),\qquad
\omega_T=2 K\left(\sqrt{r}\right),
\label{oRT}
\ee
where $K$ is the complete elliptic integral of the first kind  
and 
\be
r=\frac{s_{43} s_{21}}{s_{42} s_{31}}
\ee
is the projective-invariant ratio.

The classical string configuration is now 
given by \eq{Xcl}
with $\omega(z) $ given by \eq{SCmap} and $\omega_T$, $\omega_R$ given by
\eq{oRT}. 
The classical induced metric \rf{gcla}
is diagonal and becomes conformal  if 
\be
\frac{\omega_T}{\omega_R} \equiv
\frac{K\left( \sqrt{r}\right)}{K\left( \sqrt{1-r}\right) }= \frac{T}{R},
\label{coif}
\ee
which guarantees that the quadratic term equals the Nambu--Goto one. 
The ratio of the $K$'s in \eq{coif} is known as the
Gr\"otzsch modulus which is monotonic in $r$.

For the solution \rf{Xcl} we have for the (diagonal) induced metric
\be
\sqrt{g_{\rm cl}(x,y)}= \frac {RT}{\omega_R\omega_T} \frac{s_{42}s_{31}}
{\prod_{i=1}^4 \sqrt{(x-s_i)^2+y^2}},
\label{sqrtg}
\ee 
which becomes singular at the boundary for $x=s_i$.
We thus regularize by moving the boundary from the real axis
slightly into the complex plane, \ie replacing $y=0$ by $y=\eps(s)$
and denoting $\eps_i=\eps(s_i)$. The boundary metric 
\be
e(s)\equiv \sqrt[4]{g(s,\eps(s))}
\ee
has then gotten regularized:
\be
e(s_j)= \sqrt \frac {RT}{\omega_R\omega_T} \frac{s_{42}s_{31}}
{\prod_{i\neq j} \sqrt{|s_j-s_i|}\eps_j}.
\ee
On the other hand 
\be
\eps(s)=\eps /e(s)
\label{epss}
\ee
and
\be
\eps_i=\eps /e(s_i),
\label{epsi}
\ee
where $\eps$ is a physical cutoff of the dimension of length,
for general covariance,
as was pointed out in \cite{Polyakov-1981}. We then obtain
\bea
\eps_1&=&\eps^2 \frac {\omega_R\omega_T}{RT}\frac{s_{41}s_{21}}{s_{42}},\non
\eps_2&=&\eps^2 \frac {\omega_R\omega_T}{RT}\frac{s_{32}s_{21}}{s_{31}},\non
\eps_3&=&\eps^2 \frac {\omega_R\omega_T}{RT}\frac{s_{43}s_{32}}{s_{42}},\non
\eps_4&=&\eps^2 \frac {\omega_R\omega_T}{RT}\frac{s_{43}s_{41}}{s_{31}}.
\label{epsis}
\eea

While \eq{epss} is quite general, it was understood in \cite{DOP84}
that for a rectangle it means a regularization of the geodesic 
curvature
\be
k_g(s)= -\frac 1{2 e(s)} \partial_y \log \sqrt{g}\Big|_{y=\eps/e(s)} 
\ee  
 at the corners. For thus smeared rectangle we have 
\be
k_g(s_i) = \frac1{2\eps},
\ee
regularizing divergences at the corners in a nice way. 

\subsection{Boundary action\label{ss:2.3}}

If a function is harmonic in a simple connected domain $D$, it is 
uniquely determined by its values on the boundary $\prt D$  by the 
general Poisson formula. This formula comes simple and useful 
if we map $D$ to the upper half-plane such that we have a 
harmonic function $f(x,y)$ given by its value $f(s,0)$
at the boundary. We now apply this to a solution $X_\mu^{\psi}(z)$
of \rf{ja4}
and thus rewrite the first term on the right-hand side of \eq{ja5}
as a boundary contribution
\bea
\lefteqn{\frac 12 \int d^2 z \, \partial_a X^\psi  \cdot
\partial_a X^{\psi}}\non && = \frac 1{4\pi} \int_{-\infty}^{+\infty} \d s_1 \d s_2 
 \frac{[X^\psi(s_1,0) - X^\psi(s_2,0)]^2}{(s_1-s_2)^2}.
\label{bouc}
\eea

The boundary curve $C$ is generically described by the function $x^\mu(t)$
with a certain choice of the parameter $t$.
If we change the parametrization $t\to f(t)$ ($f'(t)\geq0$),
the function $x^\mu(f(t))$ will describe the same boundary curve $C$. 
As was emphasized in \cite{Polyakov-1981}, for a general curve
we have
\be
X_{\mu}^\psi(s,0) = x_\mu(t(s))
\label{Xbcx}
\ee
where the presence of 
a reparametrizing function $t(s)$ ($t'(s)\geq 0$) is required 
at the classical level to fulfill the conformal gauge 
(or the Virasoro constraints) and  $t(s)$ depends upon the curve $C$.
Equation~\rf{ja3} is of this type for the proper-length parametrization.
$t(s)$ is related to the boundary value $\psi(s)$ of the Liouville field as
\be
t'(s)=\e^{\psi(s)/2-\psi(t(s))/2}
\ee
(notice that $\psi(s)=0$ for the proper-length parametrization). 
At the quantum level we have to path integrate over $t(s)$ \cite{Alv83}, 
which is required for the consistency of 
the Polyakov string formulation.
It is basically the same as the path-integration over $\psi(z)$ in
\eq{ja5}.

Substituting \eq{Xbcx} into the right-hand side of \eq{bouc},
we can rewrite  the boundary action as Douglas' integral 
\be
S_{\rm B}=\frac 1{4\pi} \int_{-\infty}^{+\infty} \d s_1 \d s_2 
 \frac{[x(t(s_1))-x(t(s_2))]^2}{(s_1-s_2)^2},
\label{DI}
\ee 
whose minimization with respect to $t(s)$ determines the minimal
surface spanned by the curve $C$ (or the classical string configuration
$X_{\mu}^\psi(z)$), i.e.\ the choice of $\psi(s)$ which ensures that 
also \rf{ja4a} is satisfied.

The classical boundary action for the rectangle can be evaluated by
substituting Eqs.\ \rf{Xcl} and \rf{oRT} into \eq{bouc}. We then find
\bea
\lefteqn{\frac 12 \int d^2 z \, \partial_a X^\psi_{\mu}  \cdot
\partial_a X^\psi_{\mu}}\non && = 
\frac12 \left[T^2 \frac{K(\sqrt{1-r})}{K(\sqrt{r})}+
R^2 \frac{K(\sqrt{r})}{K(\sqrt{1-r})}\right]
\label{Sb}
\eea
and, since at the classical level $r$ is linked to $T/R$ by \eq{coif},
we recover the area $RT$. However, if we want to perform the functional
integrations over reparametrizations of the boundary we cannot assume 
\eq{coif}  and $r$ will become an integration parameter as will be 
discussed below.

\subsection{The one-loop calculation}

We now have to perform the functional integration over 
$\phi$ for a fixed $\psi$ and then over $\psi$ in \rf{ja5}.
In performing the integral over $\psi$ we will expand 
to quadratic approximation around the parametrization 
which lead to \rf{Sb} with $r$ as a free parameter.
Since the integration of $\phi$ represents the same
order as that over $\psi$, it is consistent to keep 
the boundary condition of $\phi$ also to be the one
which leads to \rf{Sb}, i.e.\ to be independent 
of the quadratic fluctuations of $\psi$. The two calculations 
have already been performed, the $\phi$ integration in \cite{DOP84} and the 
$\psi$ integral in \cite{MO10a}. Here we basically combine the two 
results and for consistency add a few details. First we address the 
$\phi$ integral, next the $\psi$ integral.

\subsubsection{L\"uscher term in UHP coordinates\label{ss:34}}


In the world sheet parametrization one naturally expands around 
$\phi=0$.%
\footnote{The derivation of the L\"uscher term for the Polyakov string in 
the world sheet parametrization was given in Ref.~\cite{JM93}.} 
Mapping the $\omega \in \omega_R\times \omega_T$ to 
the upper half-plane we then expand around 
\be
\phi_{\rm cl}=2\log \left| \frac{\partial \omega(z)}{\partial z} \right|,
\ee
where the map $\om(z)$ was explicitly given above. We thus 
split the Liouville field into the classical
and quantum parts: $\phi=\phi_{\rm cl}+\phi_{\rm q}$. 
Then we notice that $\phi_{\rm q}$ vanishes at the boundary since the
boundary condition is already satisfied by $\phi_{\rm cl}$. The path
integral over $\phi_{\rm q}$ decouples in the expansion about
the saddle point to the quadratic approximation and shifts the 
dimension by 1 just as for the closed string \cite{KPZ,DK88,Dav89}.
The $\phi$ part of the integration in \rf{ja5} can thus,
to quadratic order, be written as 
\beq\label{ja7}
\int \cD \phi(z) \;
\e^{\frac{d-26}{96\pi} \int d^2 z \,
\partial_a \phi\partial_a \phi\,} = 
\e^{\frac{d-25}{96\pi} \int d^2 z \,
\partial_a \phi_{\rm cl}\partial_a \phi_{\rm cl}\,}.
\eeq
This shift seems unavoidable when using the Polyakov formulation.

The computation of the integral 
\be
\frac{1}{24\pi} \int d^2 z\, 
\partial_a \log \left| \frac{\partial \omega(z)}{\partial z} \right|
\partial_a \log \left| \frac{\partial \omega(z)}{\partial z} \right|,
\label{Lt}
\ee
simplifies, if we set
$s_1=0$, $s_2=r$ ($0<r<1$), $s_3=1$ and $s_4=\infty$ in the 
expression for $\om(z)$. Then
\be
\omega(z)= \int_{r}^z \frac{\d s } 
{\sqrt{(1-s)(s-r)s}}
\label{SCmap3}
\ee
and
\bea
\sqrt{g_{\rm cl}(x,y)}&= &\frac {RT}{4 K\left( \sqrt{r}\right) 
K\left( \sqrt{1-r}\right)} \non &\times&\frac{1}
{\sqrt{((x-1)^2+y^2)((x-r)^2+y^2)(x^2+y^2)}}. \non &&
\label{sqrtg3}
\eea
Here we have used \eq{oRT} for $\omega_R$ and $\omega_T$.

Modulo an inessential (logarithmic) divergence at large $y$ we find
that \rf{Lt} is given by
\bea
\lefteqn{\hspace*{-.6cm}
-\frac{1}{96} \log\left[r^2 (1-r)^2 \eps_1 \eps_2 \eps_3
  \right]+{\rm const.}}\non &=&-\frac{1}{24} \log\left[r (1-r) \right]
+{\rm const.}\,,
\label{lll}
\eea
where we have substituted \eq{epsis} for $\eps_i$'s.
Using the asymptotes
\be
K\left(\sqrt{r}\right)\stackrel{r\to1}\to \frac12 \log\frac{16}{1-r},
\qquad K\left(\sqrt{1-r}\right)\stackrel{r\to1}\to \frac\pi2, 
\label{asymptotes}
\ee
we reproduce the standard L\"uscher term (per one degree of freedom)
\be
\frac {1}{24} \log \frac{16}{1-r}=\frac{\pi \omega_T}{24 \omega_R}  
\label{TRvsst}
\ee
for $\omega_T\gg \omega_R$.\footnote{
In deriving \eq{lll} we have not assumed that $r$ is near 0 or 1, so it
can be arbitrary. In world sheet coordinates we are expanding 
around $\phi=0$ and thus simply calculating the determinant of a 
Laplacian. This determinant is a   product over the eigen modes
of the Laplacian and  
can be expressed \cite{DF83} through the Dedekind 
$\eta$-function:
\be
\prod_{m,n=1}^\infty \left( \frac{\pi m^2}{\omega_T^2}+ 
\frac{\pi n^2}{\omega_R^2} \right) = \frac1{\sqrt{2\omega_R} }\,
\eta \left(i \frac{\omega_T}{\omega_R}   \right).
\nonumber
\ee
We have numerically verified with Mathematica that indeed
\be
~~\frac1{2\left(K\left( \sqrt{1-r}\right) \right)^{1/2}} \,
\eta \left( i \frac{K\left( \sqrt{1-r}\right) }{K\left( \sqrt{r}\right) }  
 \right)= 
\frac{1}{2^{5/6} \pi^{1/2} }  \left[r(1-r)\right]^{1/12}
\nonumber
\ee
in the whole range $0\leq r \leq 1$. 
It may be interesting to know whether this identity is known to
the Math community.}

While the result~\rf{lll} is derived for the particular regularization
introduced in Subsect.~\ref{ss:2.2}, we believe it is universal,
as the L\"uscher term is.

\subsubsection{The boundary reparametrization path integral}

We finally have to perform the integral over quadratic 
fluctuations of the boundary field $\psi$ in \rf{ja5}
(the main reason we introduced the UHP). 
As was shown in \cite{MO10a}, accounting for reparametrizations in a quadratic
approximation (justified by long strings) results for an outstretched rectangle
in \eq{TRvsst} with the extra factor of 24.
The computation is sketched in Appendix~\ref{appA}. 
Adding the contribution to the effective
action from the $\phi$ and the $\psi$ integrations in \rf{ja5} 
leads to the L\"uscher term
\be
\left(1+\frac{d-25}{24} \right) \log \frac1{1-r} =
\frac{d_\perp}{24}  \log \frac1{1-r}
\label{2nd}
\ee
with $d_\perp=d-1$.%
\footnote{For the Polyakov string formulation the number $d_\perp$ of fluctuating 
transverse (physical)
degrees of freedom of the string, which the L\"uscher term is 
proportional to, is $d-1$ rather than $d-2$ because the dimension 
is effectively shifted by 1, as is already mentioned,
since the Liouville field also fluctuates~\cite{KPZ,DK88,Dav89}.
 We have $d_\perp=2$ for the QCD string.}

A subtlety in performing the integral over the $\psi$ field
is that we have to fix the 
projective $PSL(2,\Bbb{R})$ symmetry which is inherited by the  Douglas'
integral \rf{DI}. Technically, it is convenient to split 
the group of reparametrizations of the rectangle onto 
reparametrizations at the edges of the rectangle with fixed 
corners, that is to fix
\be
t(s_i)=s_i \qquad \hbox{for}~ i=1,2,3,4 .
\ee
These reparametrizations simply relabel the points of the same edge.
Thanks to the projective symmetry, the result depends on
only one parameter: the ratio $r$. 
Additionally, we have to reparametrize at four corners,
that is to integrate over four $s_i$'s, leaving their order at the real axis.
This gives the (infinite) volume of  $PSL(2,\Bbb{R})$ times
the integral over $r$, which in the semiclassical (or one-loop)
approximation is saturated by the classical value
determined by  \eq{coif}. 

Adding \rf{Sb} and \rf{2nd} we finally obtain for $(1-r)\ll1$
the following effective action
\be
S_{\rm eff}= \frac1{4\pi\alpha'} \left[T^2 \frac{\pi}{\log{\frac1{1-r}}}+
R^2 \frac{\log{\frac1{1-r}}}{\pi} \right]-\frac{d_\perp}{24}  \log \frac1{1-r}.
\label{Sefff}
\ee
We emphasize again that the formula does not rely on  $r$
obeying  \eq{coif} (but it assumes $(1-r)\ll1$). 
For $r$ not satisfying \eq{coif} 
the induced metric tensor is diagonal but not conformal (or isothermal).

Beyond the semiclassical approximation we have to integrate the
exponential of (minus) \rf{Sefff} over $r$ from 0 to 1. There are two
important cases, where this integral has a saddle point:
$T \gg R$ and large $d$.
Then we can simply replace the integration by
the minimization with respect to $r$. Technically, this reminds
the derivation of the Regge limit of the Veneziano amplitude.
Such an analogy is considered in detail in Ref.~\cite{Mak11b}.

The limit of large $|d|$ is generically the one, where a mean-field 
approximation becomes exact. We can therefore look at $r$ as if it was
a parameter of the variational mean-field ansatz. Its best
description of an exact result is reached at the minimum.
For $|d|\to\infty$ the minimum gives the exact energy of the
string ground state, while for finite $d$ we may generically expect 
an approximate value larger than the exact one. However, it
can be argued  for the Polyakov formulation of the Nambu--Goto string that
we may expect to obtain the exact result for
$T\gg R$ even at finite $d$ by the mean field.

\subsection{Spectrum of pure Nambu--Goto}

To compute the spectrum, we have to minimize, as is shown in the previous
Subsection, 
the effective action \rf{Sefff}
with respect to $r$. The saddle-point equation is quadratic in $\log(1-r)$
and has the solution
\be
 \frac{\log{\frac1{1-r_*}}}{\pi}= 
\frac{T}{ \sqrt{R^2-R_0^2}},
\label{r*}
\ee
where $R_0^2$ is the inverse tachyon mass squared
\begin{eqnarray}
R_0^2&=&\frac{\pi^2 d_\perp}{6} \alpha' \qquad ~\,\hbox{for open string}, \non
R_0^2&=&\frac{2\pi^2 d_\perp}3 \alpha' \qquad \hbox{for closed string}
\end{eqnarray}
with the periodic boundary condition%
\footnote{The L\"uscher term is then four times larger than for
an open string.}
 along the 1-direction.
This replaces the pure classical \eq{coif}.
At the minimum we have
\be
S_{\rm eff}= \frac T{2\pi\alpha'} \sqrt{R^2-R^2_0},
\label{A-A}
\ee
reproducing the Alvarez--Arvis energy 
\cite{Alv81,Arv83} of the ground state. Our derivation
extends to UHP that of \cite{Mak11b} given for the world sheet coordinates.

For the average of the induced metric tensor we have \eq{mmm} with
\bea
\rho&= &
 \frac{\bar \rho}
{\sqrt{((x-1)^2+y^2)((x-r)^2+y^2)(x^2+y^2)}}, \non \bar \rho&=&
\frac{R^2-R_0^2/2}{\omega_R^2}  ,
\label{rhorho}
\eea
which is fulfilled owing to \eq{r*}. It replaces the classical
\eq{sqrtg3}, which is recovered as $\alpha'\to0$.
For the ratio of the area of a typical surface to the minimal area
we then find
\be
\frac{\left\langle A \right\rangle_\phi}{RT}=
\frac{R^2-R_0^2/2}{R\sqrt{R^2-R^2_0}}.
\label{avA}
\ee
It tends to 1 as $R\to\infty$, but blows up near the tachyonic singularity 
at $R=R_0$, when crumpling of surfaces occurs. Then we can no longer trust in
the mean-field approximation.

\section{String with extrinsic curvature\label{s:3}}

A simplest generalization of the  Nambu--Goto string is
bosonic string with an extrinsic curvature, which is
the next-order operator after the Nambu--Goto action.
It was first introduced for QCD string 
in~\cite{Pol86,Klei86}. 
The original idea was that it 
provides rigidity of the string that makes it smoother.
The spectrum of 
such a rigid string is investigated in~\cite{OY86,Bra86,GK89} 
(for a review see Ref.~\cite{Ger91}). We briefly review 
some of these results below.

The action of the bosonic (Polyakov) string with  the extrinsic curvature term 
reads
\be
S_{\rm r.s.} = \frac {\K}2 \int d^2 z\, 
\sqrt{g} g^{ab}\partial_a X\cdot
\partial_b X+ \frac{1}{2\alpha} 
\int d^2 z \,{\sqrt{g}} \Delta X\cdot \Delta X,
\label{r.s.}
\ee
where $K=1/2\pi \alpha'$ is the string tension,
${\alpha}$ is {a dimensionless} constant (rigidity)
and $\Delta$ is the 2d Laplace--Beltrami operator.

It is custom to consider $g_{ab}$ as an induced metric and, 
introducing $\rho=\sqrt{g}$ and the Lagrange multipliers $\lambda^{ab}$,
to rewrite the action ~\rf{r.s.} in the conformal gauge as
\bea
S_{\rm r.s.} &= &\K \int d^2 z\, \rho + \frac{1}{2\alpha} 
\int d^2 z \,\frac1{\rho}\, \partial^2 X\cdot \partial^2 X 
 \non && +\frac12 \int d^2 z \,\lambda^{ab}
\left(
\partial_a X \cdot\partial_b X - \rho \delta_{ab}
\right). 
\eea

\subsection{World sheet parametrization\label{ss:ws}}

Likewise for the Nambu--Goto string, we consider the world sheet parametrization
and restrict ourselves with
a mean-field (variational) ansatz, when only $X^\perp$ fluctuates. 
It is exact at large $d$ for $\K\sim d$ and $\alpha\sim 1/d$.
We write (for ${\omega_T}=T$)
\begin{eqnarray}
&&X^1_{\rm mf}(\omega)=\frac {\omega_1}{\omega_R}{R}, \quad
X^2_{\rm mf}(\omega)=\omega_2, \quad
X^\perp(\omega)=\delta X^\perp(\omega),  \non 
&&\rho_{\rm mf}(\omega)= \rho, \quad 
\lambda_{\rm mf}^{11}(\omega)= \lambda^{11},\quad
\lambda_{\rm mf}^{22} (\omega)= \lambda^{22}, \non &&
\lambda_{\rm mf}^{12} (\omega)= \lambda^{21}_{\rm mf}(\omega)=0 
\end{eqnarray}
and
\begin{eqnarray}
\lefteqn{\frac 1T S_{\rm mf}}\non
&=&\frac 12 \left( \lambda^{11} \omega_R + \lambda^{22}
\frac {R^2}{\omega_R} \right) +
\rho\left(\K -\frac{\lambda^{11}}2-\frac{\lambda^{22}}2 \right)\omega_R \non
&&+\frac{d}{2T} \,{\rm tr}\, 
\log \left( -\frac{\lambda^{11}}\rho \partial_1^2-
\frac{\lambda^{22}}\rho \partial_2^2 
+\frac{1}{{\alpha}\rho^2} (\partial_1^2+\partial_2^2)^2 \right). \non &&
\label{tttd}
\end{eqnarray}

The determinant in the last line of \eq{tttd} can be evaluated using
the formulas of Refs.~\cite{OY86,Bra86,GK89}.
Using a momentum-space representation,
integrating 
over $\d k_2$ (as $T \to \infty$), regularizing
via {the zeta function}
and introducing
\be
\Xi = \frac {\sqrt{{\alpha} \rho \lambda^{11}}\,\omega_R}{2\pi}
\ee
{instead of $\rho$}, we get 

\pagebreak
\begin{widetext}
\begin{eqnarray}
\frac 1T S_{\rm mf}&=&\frac 12 \left( \lambda^{11} \omega_R + \lambda^{22}
\frac {R^2}{\omega_R} \right) +
\left(\frac{2\K(\mu)}{\lambda^{11}} -1-\frac{\lambda^{22}}{\lambda^{11}}
 \right) \frac{2\pi^2 \Xi^2}{{\alpha(\mu)}\omega_R } 
\non &&+ \frac{2\pi d }{\omega_R}\left\lbrace -\frac16+\frac{\Xi}2 
+\sum_{n\geq1} \left[
\sqrt{\frac{\Xi^2}2+n^2
+\Xi\sqrt{\frac{\Xi^2}4+
\left(1-\frac{\lambda^{22}}{\lambda^{11}}\right)n^2}}\right.\right.
\non
&&~~~~~~~~~
\left.\left.
+\sqrt{\frac{\Xi^2}2+n^2
-\Xi\sqrt{\frac{\Xi^2}4+
\left(1-\frac{\lambda^{22}}{\lambda^{11}}\right)n^2}}
-2n -\frac{\Xi^2}{4n}\left(1+\frac{\lambda^{22}}{\lambda^{11}}\right)
\right]\right\rbrace. 
\label{26}
\end{eqnarray}
\end{widetext}

In \eq{26} we have  introduced  an ultraviolet cutoff $a_{\rm UV}$ by
\be
\sum_{n\geq 1} \frac 1n=\log \frac1{\mu a_{\rm UV}}.
\ee
and performed the renormalization of the parameters 
$K$ and $\alpha$ of the bare action 
by introducing (renormalized)
\be
{\alpha}(\mu)= 
\frac{{\alpha}}{1-\displaystyle{\frac{{\alpha}d}{4\pi}
\log \frac{1}{\mu a_{\rm UV}}} }\,,\qquad
\K(\mu)= \K 
\frac{\alpha(\mu)}{\alpha }
\ee
as is prescribed by asymptotic freedom of the model~\cite{Pol86,Klei86}.
Then UV divergences disappear in \eq{26} and the result is finite.

As $R\to\infty$ the action~\rf{26} can be minimized iteratively,
order by order in $\alpha'/R^2$,
which is nothing but a semiclassical expansion, starting from the classical
approximation:
\be
\omega_R=R, \quad \lambda^{11}=\lambda^{22}=K, \quad
\rho=1.
\ee
Because $\lambda^{11}=\lambda^{22}$ order by order of the
expansion in $\alpha'/R^2$ or, correspondingly, in $1/\Xi^2$,
the sum over $n$ in \eq{26} is exponentially suppressed~\cite{GK89}.
We thus reproduce the same solution as for the Nambu--Goto string to any order
in $\alpha'/R^2$.

However, the extrinsic curvature becomes important at 
$\alpha'/R^2\sim1$. The case of small $\alpha$ can be analyzed 
analytically. Then the extrinsic curvature dominates over the Nambu--Goto 
term and great simplifications occur in \eq{26}.
For small $\alpha$ we have $\Xi$ also small
and the determinant in the last line in \eq{tttd} equals
\be
\frac{d}{2T} \,{\rm tr}\, 
\log \left( \ldots \right) \longrightarrow  
 -\frac{\pi d}{3 \omega_R}+
\frac d2 \sqrt{{\alpha}\rho\lambda^{11}} 
 \quad \hbox{({closed string})}
\ee
as  $\alpha\to0$. 
The first term on the right-hand side is twice larger than usual L\"uscher's
term,
because the quartic operator dominates over quadratic that doubles
degrees of freedom. The second term comes from zero modes.

For small $\alpha$ we can easily minimize the mean-field
action~\rf{tttd} to obtain~\cite{PY92}
\bea
E_0&= &\lambda^{11}\omega_R, \non \sqrt{\lambda^{11}}&=&
\frac 38 \frac{d\sqrt{{\alpha}}}R 
+\sqrt{\frac 9{64}\frac{d^2 {\alpha}}{R^2}+
\K-\frac{\pi d}{3 R^2}},\non
\omega_R&=&\sqrt{R^2-\frac{dR}2 
\sqrt{\frac{{\alpha}}{\lambda^{11}}}}
\label{gse}
\eea
and
\be
\rho=\frac {R^2}{\omega_R^2}.
\ee
These formulas are applicable for small $\alpha$ if
\be
R^2 \ll \frac{\alpha'}{\alpha}
\label{dddo}
\ee
when $\Xi$ is small. 

\subsection{UHP parametrization}

With the extrinsic-curvature term included, the contribution of
quantum fluctuations to the effective action is given by 
 the determinant in \eq{tttd}
which is not computable by the Seeley expansion 
like in the massive case (see Appendix~\ref{ss:Seeley} 
for a review). It
can be easily evaluated, however, for either large or small $\alpha$.
For large $\alpha$ we can neglect the quartic operator, reproducing
the above results for the conformal anomaly. For small $\alpha$
we can instead neglect the quadratic operator, so the result
is simply twice larger than computed above. This factor of 2 is
simply related with doubling of degrees of freedom in the conformal
anomaly, which produces twice larger L\"uscher's term.

We thus obtain at large $d$
\bea
\frac{d}{2T} \,{\rm tr}\, 
\log \left( -\frac{\lambda^{11}}\rho \partial_1^2-
\frac{\lambda^{22}}\rho \partial_2^2 
+\frac{1}{{\alpha}\rho^2} (\partial_1^2+\partial_2^2)^2 \right)\non
= \left\{  
\begin{array}{ll}
\displaystyle{-\frac{\pi d }{6\omega_R}\sqrt{\frac {\lambda^{22}}{\lambda^{11}}} }&\quad \alpha\to \infty\\ \mbox{}&\quad \\
\displaystyle{-\frac{\pi d }{3\omega_R}}&\quad \alpha\to 0
\end{array}     
\right. .
\eea

The saddle-point minimization in the case $\alpha\to \infty$ is like as in
Ref.~\cite{Alv81} and gives [cf.\ \eq{rhorho}]
\be
\lambda^{11}=\lambda^{22}=\K,\quad \omega_R=\sqrt{R^2-\frac{\pi d}{3\K}},\quad
\bar\rho=\frac{\displaystyle{1-\frac{\pi d}{6K R^2}}}
{\displaystyle{1-\frac{\pi d}{3K R^2}}},
\ee
reproducing \eq{A-A}.

The saddle-point minimization in the case $\alpha\to 0$ is like as in
Ref.~\cite{PY92} and gives
\be
\lambda^{11}=\K-\frac{\pi d}{3R^2},\quad
\lambda^{22}=\K+\frac{\pi d}{3R^2},\quad \omega_R=R,\quad
\bar \rho=1.
\label{la2}
\ee
For the energy of the ground state we obtain
\be
E_0(R)=K R -\frac{\pi d}{3R}
\label{eee0}
\ee
to be compared with \eq{A-A} for the iterative solution.
We see that the semiclassical one-loop result
is exact as $\alpha\to0$.
This saddle point is simply the $\alpha\to0$ limit of \rf{gse}.

The domain \rf{dddo}, where Eqs.~\rf{la2}, \rf{eee0} are applicable, 
overlaps with 
the domain $\alpha'\ll R^2$, where the iterative solution coincides with the
Nambu--Goto one modulo exponential terms. However, the ground state energy
\rf{eee0} is smaller than \rf{A-A}, 
so it is the solution \rf{la2} that is realized for  
$R^2 \ll {\alpha'}/{\alpha}$.
We emphasize once again that
\eq{eee0} applies {\it only}\/ in the domain \rf{dddo}. 
For large values of $R^2\ga 1/K\alpha$ 
the iterative solution  \rf{A-A} applies.

From Eqs.~\rf{la2} we deduce that
\be
\frac{\langle A \rangle_\phi}{RT}=1
\ee
as $\alpha\to0$ in contrast to \eq{avA}. We see that the string indeed
becomes rigid and behaves as a stick as $\alpha\to0$. 
Because the ground-state energy \rf{eee0} is lower than \rf{A-A},
the string behaves as rigid for $R^2\ll 1/\K\alpha$ and as the usual
Nambu--Goto string for $R^2\gg 1/\K\alpha$. The change between the regimes 
\rf{A-A} and \rf{eee0} cannot be seen, therefore, within the
$1/R$-expansion. 

The fact that the dependence \rf{eee0} is not seen in the rather precise
lattice Monte-Carlo spectrum~\cite{ABT10} 
(see \cite{LP12} for a review and references therein) 
probably means that a large rigidity term is ruled out for QCD string.

\section{External geometry of effective string\label{s:33}}

Trajectories $X^\mu(z)$ of strings moving in $d$-dimensional 
space are  surfaces embedded in Euclidean $R^d$ space. Therefore, the  
differential geometry of the string world sheet is totally defined by the 
induced metric 
$g_{ab}=\partial_a X_\mu\partial_b X^\mu$ and second quadratic forms  
$h^j_{ab}={n^j}_\mu\,\nabla_a\partial_bX ^\mu$
and $ H^{ij}_a={n}^i _\mu\partial_a { n}^{j,\mu} $,
where $j=1,\ldots, d-2$ counts  coordinates transversal to the surface. 
For immersed surfaces, which means that the induced metric
$ g_{ab}$ is not singular, the first and second fundamental forms should 
fulfill the Peterson--Codazzi equations
\bea
 \label{peterson}
{\cal R}_{dacb}&=&h^{j}_{ab} h^{j}_{cd}-h^{j}_{ac} h^{j}_{bd},\nn\\
\epsilon^{ab} \partial_a H_b^{ij}&=& 
\epsilon^{ab} H_a^{ik} H_b^{kj}+\epsilon^{ab}h^{i,c}_a h^{j}_{b c},\nn\\
\epsilon^{ab} \nabla_a h^{i}_{bc}&=&\epsilon^{ab} h_{ac}^j H_{b}^{ji},
\eea
where the summation over repeating indexes $ i,j, a, b,\ldots$ is implied.
These equations take place because the
tangent vectors $\partial_a X^\mu,\;a=1,2$ are derivatives and together with 
the normal vectors ${n}_\mu^j$ 
they form a complete
set of basic vectors in the target space of embedding, namely 
$\partial_a X^{\nu} \partial^a X^{\nu'}+n^{j,\nu}n^{j,\nu'}=\delta^{\nu \nu'}$. 
In two  dimension, due to the symmetry properties of the 4-range curvature 
tensor 
${\cal R}$ over its indexes,
all the information reduces to the scalar Gaussian curvature $R$
\bea
\label{GC}
{\cal R}_{dacb}&=& \frac{R}{2}(g_{ab}g_{dc}-g_{ad}g_{bc}).
\eea

Therefore, we can reduce Gaussian curvature
$R$ to the second fundamental h-forms. 
Moreover, due to the definitions of the fundamental forms 
$g_{ab},\; h_{ab}^j,\; H_a^{ij}$
 and the Peterson-Codazzi equations,
 all scalars of the  surface coordinates $X^\mu$, as well as 
their higher covariant derivatives, can be always reduced to scalar 
polynomials of the second fundamental forms $h$ and $H$. 

We suggest that the dynamics of strings  
is defined by a quadratic elliptic operator ${\cal D}$, which may explicitly 
depend on external geometry
of surfaces and is reparametrization invariant. 

As it follows from the work \cite{Polyakov-1981}, any  2d gravity theory 
will always have a quantum anomaly
of conformal transformations $g_{ab}\rightarrow \rho g_{ab}$, unless some 
cancellation 
between bosons, fermions and ghosts will take place at certain critical 
dimensions. 
The anomaly expresses itself in the fact that, though 
the trace of the stress-energy tensor
is zero  at the classical level, its quantum average is nonzero 
and is connected with the expansion of the heat-kernel operator
$e^{- {\cal D}/{\mu^2}}$, where $\mu^2$ is a regularization scale 
of quantum fluctuations.   
Following the works \cite{Polyakov-1981, Olesen}, we 
expect that each central charge 1 will give the following contribution 
to the trace of the stress-energy tensor
\bea \label{anomaly}
\lefteqn{\langle T_a^a \rangle = \langle \e^{-{\cal D}/{\mu^2}}\rangle
=\frac{1}{24 \pi} \sqrt{g} R(\xi_a)+ \frac{c_0}{16 \pi}  \sqrt{g} (h^j)^2 }
\non &&
+ \mu^2 \sqrt{g}\nn+ \mu^2 \sqrt{g} 
F\big[\frac{1}{\mu^2} h^i_{ab}K\big[\frac{1}{\mu^2}H^2\big]^{ij}h^j_{cd} \big].
\non &&
\eea
Here $g=\det[g_{ab}]$,  $R(\xi_a)$ and $h^j= g^{ab}h^j_{ab}$ ($j=1,\ldots, d-2 $)  
are the  Gaussian and
mean sectional curvatures, respectively,  and
\be
\label{K}
K\big[\frac{1}{\mu^2} H^2\big]=
1+\frac{k_2}{\mu^2}H_{s_1} H_{s_2}+\ldots+ 
\frac{k_{2m}}{\mu^{2m}}H_{s_1}\cdots H_{s_{2m}}\cdots
\ee
is an even-order series of the matrices $H^{ij}_s$, while 
$F[\frac{1}{\mu^2} h^j_{ab}h^j_{cd} ] $ is a scalar 
re\-parametriza\-tion-invariant series of its 
tensor arguments $\frac{1}{\mu^2} h^j_{ab}h^j_{cd} $, which starts from a square. 
 Symbolically, we can write
\be
\label{seeley}
F\big[\frac{1}{\mu^2} h^j_{ab}h^j_{cd} \big]= 
P_2\big[\frac{1}{\mu^2} h^j_{ab}h^j_{cd} \big]^2+
P_3\big[\frac{1}{\mu^2} h^j_{ab}h^j_{cd} \big]^3
+\ldots
\ee 
The notations $P_k$, $k=2,3,\ldots$, in this expression mean scalar polynomials
 of k'th order of homogeneity, which are obtained
by all type of contractions of the indexes in 
$ h^{j_1}_{{a_1}{b_1}}h^{j_1}_{{c_1}{d_1}}\otimes \cdots \otimes  
h^{j_k}_{{a_k}{b_k}}h^{j_k}_{{c_k}{d_k}} $ by using the metric $g^{ab}$.
As we see below, the terms $\sum_{j=1}^{d-2}h^j_{ab}h^j_{cd}$ produce
the coefficients of $d-2$, while an insertion of $H^{ij}$
does not change it, but increases the order of $1/\mu^2$. 
Therefore, we should drop
all $H^{ij} $'s  to the leading order in ${(d-2)}/{\mu^2}$ 
and take $K\big[\frac{1}{\mu^2} H^2\big]=1$.

The expansion of $\langle \e^{-{\cal D}/{\mu^2}}\rangle$, that enters \eq{anomaly}, 
in the powers of $\mu^{-2 k}$ is known as the Seeley 
expansion. It is a standard way of calculations 
for quadratic elliptic operators ${\cal D}$ as presented in 
Refs.~\cite{Gil75,Olesen,Alv83}.
The first term in the expression (\ref{anomaly}) is usual and defined by 
internal geometry of the string world sheet (\ie the metric), 
while the second term comes from external geometric characteristics 
of the surfaces. 
For example, for the Green--Schwartz superstring with N=1 supersymmetry 
 it was obtained that $c_0=1$ \cite{Kavalov-1986}. 
These terms are universal and their coefficients are dimensionless. 
Other terms have dimensionfull coefficients
and therefore break conformal invariance explicitly. 

Equation (\ref{anomaly}) for the conformal anomaly allows us to find out 
the most general form of the effective
action ${\cal L}_{eff}$. Namely, we should integrate the Word identity
\bea
\label{W-eq} 
\delta W =- \frac{\delta \rho}{2 \rho} \langle T_a^a \rangle,
\eea
where $\langle T_a^a \rangle$ in (\ref{anomaly}) should be written in 
a conformal gauge with the metric  $g_{ab}=\rho \delta_{ab}$.
In order to do that, let us first express  $ h^j_{ab}h^j_{cd} $ as a sum 
of the tensors
\bea
\label{hh}
h^j_{ab}h^j_{cd}=\frac{1}{2}h^j_{a\{b}h^j_{c\}d}-\frac{1}{2} {\cal R}_{adbc},
\eea 
symmetrized and antisymmetrized over the indexes $b,\;c$. Here the
curly brackets $\{ \}$ around the $b, c$ indexes mean a symmetrization 
over them. 
Substituting this representation into the expansion (\ref{seeley}) and using 
(\ref{GC}), we obtain
\bea
\label{Q}
F\big[\frac{1}{\mu^2} h^j_{ab}h^j_{cd} \big]&=&
\tilde{P}_2\big[\frac{1}{\mu^2} h^j_{a\{b}h^j_{c\}d} \big]^2 
Q_2\big[\sqrt{g}R\big]\non &
+&\tilde{P}_3\big[\frac{1}{\mu^2} h^j_{a\{b}h^j_{c\}d} \big]^3 
Q_3\big[\sqrt{g}R\big]
+\cdots, ~~~~~
\eea 
where $ Q_p[R]$ are the series in the Gaussian curvature $R$
\bea
\label{QQ}
Q_p[\sqrt{g}R]=1+ w_{p1} \sqrt{g} R +\cdots + w_{ps} \sqrt{g}^s R^s+\cdots.
\eea
In the conformal gauge $R=-\frac{2}{\rho}\partial_a^2 \log[\rho]$, while
\bea
\label{h-conformal}
 h^j_{a\{b}h^j_{c\}d}=\rho \big(\partial_a {\vec n}^j {\vec e}_{\{b}\big)
\big(\partial_d {\vec n}^j{\vec e}_{c\}}\big).
\eea
A conformal variation of this expressions gives 
$\delta (\sqrt{g} R)=-2 \partial_a^{-2} \frac{\delta\rho}{\rho}$
and $\delta  h^j_{a\{b}h^j_{c\}d}= \frac{\delta\rho}{\rho}  h^j_{a\{b}h^j_{c\}d} $, 
which yields
\bea
\label{variation}
\frac{\delta\rho}{\rho}\big[\sqrt{g}R\big]^n&=&-\frac{1}{2(n+1)} 
\delta \Big[\sqrt{g}R {\vec \partial^{-2}}\big(\sqrt{g}R\big)^n\Big],\nn\\
\frac{\delta \rho}{\rho}  \Big[h^j_{a\{b}h^j_{c\}d}\Big]^n&=& 
\frac{1}{n}\delta\Big[h^j_{a\{b}h^j_{c\}d}\Big]^n .
\eea

The relations 
presented above allow us to integrate the Word identity (\ref{W-eq}) and 
to calculate the effective action
\bea
\label{W}
{\cal L}_{eff}&=& - \frac{c}{96 \pi} \sqrt{g} \big[R \Delta^{-1}R+\mu^2] 
- \frac{c_0 c}{16 \pi}\sqrt{g}\big[h^{j,a}_a\big]^2\Delta^{-1}R \nn\\
&&- c \sqrt{g}\; F_1\big[\frac{1}{\mu^2} h^j_{ab}h^j_{cd},\sqrt{g}R \big]
\non &&
-c \sqrt{g}\;\sqrt{g}R {\vec \partial^{-2}} 
F_2\big[\frac{1}{\mu^2} h^j_{ab}h^j_{cd},\sqrt{g}R \big],
\eea
where $c$ is the central charge of the system, while $F_1[\cdot]$ and 
$F_2[\cdot]$ are defined by  $F[\cdot]$ as
\bea
\label{seeley2}
F_1\big[\frac{1}{\mu^2} h^j_{ab}h^j_{cd} \big]&=& 
\frac{1}{2}\tilde{P}_2\big[\frac{1}{\mu^2} h^j_{ab}h^j_{cd} \big]^2 
Q_2\big[\sqrt{g}R\big] \non &&+
 \frac{1}{3}\tilde{P}_3\big[\frac{1}{\mu^2} h^j_{ab}h^j_{cd} \big]^3 
Q_3\big[\sqrt{g}R\big]
+\ldots,\nn\\
F_2\big[\frac{1}{\mu^2} h^j_{ab}h^j_{cd} \big]&=& 
\tilde{P}_2\big[\frac{1}{\mu^2} h^j_{ab}h^j_{cd} \big]^2 
\tilde{Q}_2\big[\sqrt{g}R\big] \non &&+
\tilde{P}_3\big[\frac{1}{\mu^2} h^j_{ab}h^j_{cd} \big]^3 
\tilde{Q}_3\big[\sqrt{g}R\big]
+\ldots
\eea
Here 
\bea
\label{Q2}
\tilde{Q}_p[\sqrt{g}R]&=&1-\frac{1}{2\times  2} w_{p1} \sqrt{g} R 
\non && +\ldots -\frac{1}{2(s+1)} w_{ps} \sqrt{g}^s R^s+\ldots
\eea


The functions $F_1[\cdot]$ and  $F_2[\cdot]$ are model dependent. 
The coefficients of their expansion in $ h^{j,a}_a\;  h^{j,b}_b $
and $R$
depend essentially on what theory we are considering: various types of 
superstrings or gauge theories.

If we have in the effective action 
additional operators of the type in \eq{W},
their correction to the spectrum can be computed by evaluating their values
for the classical solution
 \rf{Xcl}, \rf{oRT} with $r$ substituted by
$r_*$ given by \eq{r*}. This is
rigorous at least for large $d$, 
because the operator is expected to give a $1/d$
correction to the spectrum. 
Since  $h^j_{ab}=0$ and $R=0$ for the classical solution \rf{Xcl},
these additional terms in the effective action \rf{W}
do not change the spectrum.
As we have seen in the previous Section, the spectrum may change,
however, if the string length is not too large, because the first term in the
effective action \rf{W} may change.

We have ignored above in this Section possible boundary terms
in the effective action, 
which may be required by consistency, like 
the well-known Gibbons--Hawking
term in the gravitational action for an open manifold.
The boundary terms play indeed the important role for the usual
conformal anomaly (given by the first term on the right-hand side of \eq{W}),
as is already discussed, but apparently not for the other terms. 
We expect those vanish because the second fundamental form $h^j_{ab}$
vanishes for the classical solution \rf{Xcl}. 
We thus believe that the possible
boundary terms will not effect the conclusion of the previous
Paragraph, while this issue deserves a more thorough consideration.

\section{Conclusions}

We have considered QCD string as an effective string
formed by fluxes of the gluon field. Its effective action, obtained by
path integrating over short-range fluctuations, 
describes long-range stringy fluctuations.
We have used the Polyakov formulation of an open string, 
where the target-space coordinates $X_\mu$ and
the world sheet metric $g_{ab}$ are treated as independent variables. 
We have shown that the effective action has to be minimized
with respect to  $g_{ab}$ in two cases: the limit of long strings
and/or the large number  $d$ of space dimensions.
This determines the ground-state energy of the string as a function of its
length.
We have found that the spectrum of the pure Nambu--Goto string is given
by the Alvarez--Arvis formula \rf{A-A}. We have then added
the next-relevant after Nambu--Goto operator in the infrared --
the extrinsic curvature -- and have shown that the spectrum is
not changed order by order in the inverse string length, but is changed
at intermediate distances (see \eq{eee0} valid for very large rigidity). 
We have considered a most general effective action 
in the form of the conformal anomaly,
that includes external geometry, and argued
that the spectrum behaves similarly.

\subsection*{Acknowledgments}

The authors acknowledge  support by  the ERC-Advance
grant 291092, ``Exploring the Quantum Universe'' (EQU).
JA acknowledges support 
of FNU, the Free Danish Research Council, from the grant 
``quantum gravity and the role of black holes''. 
Y.~M.\ and A.~S.\ thank the Theoretical Particle Physics and Cosmology group 
at the Niels Bohr Institute for the hospitality. A.~S. acknowledges 
ARC grant 13-1C132 for partial financial support.


\appendix

\section{Reparametrization path integral: rectangle\label{appA}}

We briefly reproduce in this Appendix the computation \cite{MO10a}
of the reparametrization path integral for a rectangle, using a
different mapping of UHP onto rectangle.

Expanding  about the minimizing function $s_*(t)$:
\be
s(t)= s_*(t)+ 
\sqrt{
\frac{2 \pi\alpha'}{RT}} \,\beta (t),
\ee
we write for the action to the quadratic order in $\beta$ \cite{Rych,MO10a}:
\bea
S_2[\beta]&= &\frac{1}{4\pi \omega_R\omega_T}
\int \d s_1 \d s_2\frac{|\omega'(s_1)||\omega'(s_2)|}{(s_1-s_2)^2} \non
&& \hspace*{1.6cm}\times\left[ \beta(t_*(s_1))- \beta(t_*(s_2))\right]^2.
\label{S2}
\eea
The minimizing function $t_*(s)$ is determined by the equation
\bea
\int_r^s \frac{\d x}{\sqrt{(1-x)(x-r)x}}&=&\omega_T \frac{t}{(1-t^2)}
+\frac{\omega_R}2 \non && \qquad \hbox{for}~~r<s<1\,,  \non
\int_1^s \frac{\d x}{\sqrt{(x-1)(x-r)x}}&=&\omega_R \frac{(t^2-1)}{4t}
+\frac{\omega_T}2  \non &&\qquad \hbox{for}~~s>1\,,
\eea
so that
\be
t_*(1)=\frac{\sqrt{\omega_T^2+\omega_R^2}-\omega_T}{\omega_R}
\stackrel{\omega_T\gg \omega_R}\to \frac{\omega_R}{2\omega_T}.
\ee

It is clear that the domain $0<(s_1-1),(s_2-1)\ll1$ is essential 
in the integral in \eq{S2} for $(1-r)\ll1$. It produces a large 
contribution $\sim (1-r)^{-2}$. To see this, we expand
\be
\beta(t_*(s_2))- \beta(t_*(s_1))=\beta'(t_*(s_1)) t'_*(s_1)
{(s_2-s_1)}.
\ee
We thus obtain (disregarding $\log(1-r)$'s)
\bea
\lefteqn{S_2[\beta]} \non&= &\frac{1}{4\pi \omega_R\omega_T}
\int \d s_1 \d s_2\frac{[\beta'(t_*(s_1))t'_*(s_1)]^2 }
{\sqrt{|s_1-1||s_1-r||s_2-1||s_2-r|}}
\non &\propto& 
\int \d s_1 \d s_2\frac{[\beta'(t_*(s_1))]^2 }
{\sqrt{|s_1-1||s_1-r|}}
\frac{1}{(|s_2-1||s_2-r|]^{3/2}} 
\non
&\propto & \frac1{(1-r)^2} \int \d t \,[\beta'(t)]^2 
\label{S22}
\eea
because in this domain
\be
t'_*(s)=  \frac{4t_*(1)^2}{\omega_R \sqrt{(s-1)(s-r)}},
\ee
that determines the boundary metric. It has to be regularized by
slightly moving the boundary into UHP, as is already discussed in
Subsect.~\ref{ss:2.2}.
In passing from the first to the second line of \eq{S22}, we 
have also set $t'_*(s_1)=t'_*(s_2)$ with the given accuracy.

Computing the path integral over $\beta(t)$
by the standard mode expansion, obeying 
\be\beta(0)=
\beta(r)=\beta(1)=0
\label{ddd}
\ee 
to get rid of the projective symmetry,  we obtain
with the aid of the zeta-function regularization
\bea
\int {\cal D} \beta(t)\, e^{-S_2[\beta]} &=& \prod_{\rm modes}
\left[ r(1-r) \right]^2=\left[\frac1{\sqrt{r(1-r)}}\right]^2 \non &=&
\e^{-\log[r(1-r)]},
\eea
where we have inserted $r$ to reflect the symmetry $r\to(1-r)$ of the
rectangle. The power of 2 is because of the two sets of the modes 
which contribute to the product:
one for the interval $[0,r]$ and another for $[1,\infty]$, obeying 
the boundary condition \rf{ddd}.

This result for the reparametrization path integral
seems to be exact at large $TR/\alpha'$,
because we can then restrict ourselves by the quadratic in $\beta$
approximation. 

It might seem that the result of this Appendix for the 
reparametrization (= the boundary Liouville field) path integral
is less solid than the results for the path integral over the
bulk Liouville field of Subsect.~\ref{ss:34}, but this is not the case!
We have simply shown here how the standard results for the critical 
bosonic string can be reproduced
by the given method.

\section{Review of the Seeley-expansion method\label{ss:Seeley}}

The action, describing dynamics of the Liouville field $\phi$, 
emerges from the path
integration over $X_\mu(x,y)$ (and the ghosts)
due to ultraviolet divergences regularized
by a cutoff. For smooth $\phi$ it is the conformal anomaly displayed
in \eq{ja5}.

Generically, the contribution from  $X_\mu(x,y)$ to
the Liouville action comes from the (regularized) determinant
of the 2d Laplacian:
\be
\e^{-S_{\rm L}}=\left[\det \left( -\e^{-\phi}\partial^2 \right)\right]^{-d/2}
_{\rm Reg.}. 
\ee
For the Pauli--Villars regularization we shall consider the ratio of
determinants of the form
\be
\e^{-S_{\rm L}}=
\left[\frac{\det \left( -\e^{-\phi}\partial^2 \right)}
{\det \left( -\e^{-\phi}\partial^2 + M^2\right)}\right]^{-d/2}, 
\label{PV}
\ee
where $M$ is a regulator mass.

The standard technique for computing such determinants (reviewed in 
this Appendix) is applicable
for 
\be
\Lambda^2 \e^\phi \gg 1 , \quad\hbox{or}\quad M^2 \e^\phi \gg 1 ,
\label{ineq}
\ee 
when it results in the conformal anomaly.

The standard results for the (regularized) determinants of the 2d 
Laplacian are obtained by 
Seeley's expansion \cite{Olesen,Alv83}:
\bea
\tr\log\left( -\Delta \right)\Big|_{\rm div}
&=&-\frac 1{4\pi} \left\{
\Lambda^2\int_D-\sqrt{\pi} \Lambda \int_{\partial D} \right.
\non &&+ \left.\frac 13\log{\Lambda^2}
\left[\int_D \frac R2 +\int_{\partial D} k
\right] \right\}~~~
\label{Gilkeyd}
\eea
for the divergent part and
\be
\tr\log\left( -\Delta\right)\Big|_{\rm fin}=-\frac 1{24\pi} \left[
\int_D \frac 12 R\phi+\int_{\partial D} k\phi
\right]-\frac1{4\pi} \int_{\partial D} k  
\label{Gilkey}
\ee
for the finite part. Here
\be
k=-\frac 12 n^a \partial _a  \phi
\label{k}
\ee
is the geodesic curvature and $n^a$ is
the inward normal unit vector. 

Equation \rf{Gilkey} for the finite part can be derived as follows.
Let us apply the variational derivative ${\delta}/{\delta \phi(z)}$
to  the (regularized) determinant 
\be
\tr\log\left( -\Delta + M^2 \right)\Big|_{\rm Reg}= -
\int_{\Lambda^{-2}}^\infty \frac{\d \tau}{\tau} \,\tr \e^{\tau(\Delta-M^2)}
\label{rrrrr}
\ee
and represent the result as $\partial/\partial\tau$
plus an additional term. We then obtain
\bea
\frac{\delta}{\delta \phi(z)}
\tr\log\left( -\Delta + M^2 \right)\Big|_{\rm Reg}= 
\langle z |\e^{(\Delta-M^2)/\Lambda^2}  | z\rangle \non - M^2
\int_{\Lambda^{-2}}^\infty {\d \tau}
\langle z |\e^{\tau(\Delta-M^2)}  | z\rangle .
\label{rrrrrp}
\eea

For $M^2=0$ we can
substitute the Seeley expansion of the heat kernel in the first term on the
right-hand side \cite{Gil75,Olesen,Alv83}: 
\be
\langle z |\e^{\Delta/\Lambda^2}  | z\rangle =
\Lambda^2 a_0 +\Lambda a_1 +a_2 
\label{Seeley}
\ee
with
\bea
a_0=\frac 1{4\pi},\quad 
a_1=-\frac{1}{8\sqrt{\pi}}\, \delta^{(1)}(z-z_{\rm B}),\non
 a_2=\frac{1}{12\pi} \left( \frac 12 R + k \, \delta^{(1)}(z-z_{\rm B})
\right) 
\eea
and
reproduce the conformal anomaly on the right-hand side of \eq{Gilkey}.
For the second term we can use this expansion only for large $M^2$,
when small $\tau$ are essential in the integral, to
obtain
\bea
\lefteqn{{\frac{\delta}{\delta \phi(z)}
\tr\log\left( -\Delta + M^2 \right)\Big|_{\rm Reg} }
 \stackrel{{\rm large}\; M}
= 
\Lambda^2 a_0 +\Lambda a_1} \non &&\mbox{}  \hspace*{1cm}- M^2 a_0 
- M^2 a_0 \log \frac{\Lambda^2}{M^2}
-\frac 12 M a_1.
\label{larM}
\eea

The ratio in \eq{PV} is analogously regularized as 
\be
\tr\log\frac{ -\Delta}{\left( -\Delta + M^2 \right)}= -
\int_{1/\Lambda^2}^\infty \frac{\d \tau}{\tau} \,\tr \e^{\tau\Delta}
\left(1-\e^{-\tau M^2} \right),
\label{PV2}
\ee
which is still logarithmically divergent at small $\tau$ because the Seeley 
expansion \rf{Seeley} starts from the term proportional to $1/\tau$.
For large $M$ this is explicitly seen in \eq{larM}. 

To get rid of this logarithmic divergence, we may consider the ratio
\be
{\cal R}^{(2)}\equiv\frac{\det(-\Delta)\det(-\Delta+2M^2)}{\det(-\Delta+M^2)^2},
\label{newR}
\ee
when
\be
\tr\log{\cal R}^{(2)}= -
\int_{0}^\infty \frac{\d \tau}{\tau} \,\tr \e^{\tau\Delta}
\left(1-\e^{-\tau M^2}\right)^2
\label{PV22}
\ee
is convergent. As $M\to\infty$, we have
\be
\left(1-\e^{-\tau M^2}\right)^2 \to \Theta\left(\tau-M^{-2}\right),
\ee
where $\Theta$ is the Heaviside step function, reproducing \eq{rrrrr}
with $\Lambda=M$.

\end{document}